\documentclass[a4paper,aps,twocolumn]{revtex4}
\RequirePackage[english]{babel}
\RequirePackage[latin1]{inputenc}
\RequirePackage[T1]{fontenc}
\RequirePackage{mathrsfs}
\RequirePackage{amsmath}
\RequirePackage{amssymb}
\RequirePackage{amsbsy}
\RequirePackage{color}
\RequirePackage{bm}
\begin{document}
\title{\bf{Torsion Gravity for Dirac Particles}}
\author{Luca Fabbri}
\affiliation{INFN \& Dipartimento di Fisica, Universit\`{a} di Bologna,\\
Via Irnerio 46, 40126 Bologna, ITALY}
\date{\today}
\begin{abstract}
In this paper we consider torsion gravity in the case of the Dirac field, and by going into the rest frame we study what happens when a uniform precession as well as a phase are taken into account for the spinor field; we discuss how partially conserved axial-vector currents and torsion-spin attractive potentials justify negative Takabayashi angle and energy smaller than mass: because in this instance the module goes to zero exponentially fast then we obtain stable and localized matter distributions suitable to be regarded as a description of particles.
\end{abstract}
\maketitle
\section{Introduction}
The Riemann-Cartan geometry is the result of letting the torsion find its natural place beside the curvature of space-time \cite{C1,C2,C3,C4}; Riemann-Cartan geometry is the basis for the Einstein--Sciama-Kibble gravitation, which is the theory having new field equations coupling torsion to spin in parallel to the usual field equations coupling curvature to energy of matter \cite{S,K}: thus, in this context, both spin and energy of Dirac spinor matter fields are coupled, and hence the ESK gravity is the theory that realizes in the most extensive of manners the coupling of the geometry to its spinorial matter content \cite{h-h-k-n}. The fact that there is a torsion, and that it couples to the spin, does not specify the character of torsion, nor what type of coupling it has with the spin, and the ESK-Dirac theory as presented in the references above is only the simplest of its kind: in a more general case, DESK theories can be built either by allowing torsion to propagate, or by having it coupled to spin with different constants for each of the spinor fields, or yet both of these characters \cite{Fabbri:2014dxa}. Such extended types of DESK gravities can be reduced to the simplest Einstein gravity supplemented by one axial-vector massive boson coupling to the spin in a way that is similar to the one in which the Higgs boson couples to the fermions, and as a matter of fact the torsion-spin coupling can be regarded as the axial-vector analog of the Yukawa scalar potential.

This torsionally-induced interaction can have a variety of physical effects: in the simplest case of DESK theories, that is those for which torsion does not propagate, some applications may stretch from cosmology, where torsion is used to enforce Big-Bounce \cite{Magueijo:2012ug}, to particle physics, in which torsion is used to induce a correction for neutrino oscillations \cite{a-d-r}; in the more general instances, and that is those where torsion is permitted to propagate, there may even be a coupling to scalars allowing to study simplified axial-vector models of Dark Matter \cite{Belyaev:2016icc}. An investigation on stability configurations has also been done \cite{Fabbri:2015xga}.

We remark, however, that none of these studies has yet given a definitive answer to the question regarding what influence torsion has in the context of modern physics.

To cope with this, we restart from an unrelated issue.

One of the most subtle problems in all Quantum Field Theory is that amplitudes are computed in terms of free point-like particle solutions, represented by plane waves, which are not square-integrable \cite{p-s}; these solutions are therefore not physical, and in order to have this problem circumvented one may insist that solutions have compact support. But although compactness of the support might appear to be a reasonable working hypothesis, it is imposed by force rather than being dynamically obtained. 

Therefore the issue of finding solutions that are square-integrable is also an open problem in modern physics.

In this paper we consider the possibility that these two issues, that is finding a possible effect of torsion and solving the problem of square-integrability of solutions, may be addressed simultaneously, or in other words we will try to see if the square-integrability of the solutions can be granted by the presence of torsion: to do this we start by reviewing the most general theory of propagating torsion in gravity for Dirac spinor matter fields; we will continue the study by boosting into the frame that is at rest with respect to the overall motion but which is also rotating to follow the uniform precession and unitary phase of the spinor matter field. In such a setting, we will show that for a specific choice of the momentum and in the effective approximation, the large-distant behaviour of the square of the module is that of decreasing exponential function.

We conclude with a comparison to other approaches.
\section{Geometrical Dynamics of Matter}
In this work we follow notation and conventions of \cite{Fabbri:2014dxa}.

For all the basic notations, the metric $g_{\alpha\rho}$ will be used to move coordinate (Greek) indices; tetrads $e^{\alpha}_{a}$ are taken to be ortho-normal $g_{\alpha\rho} e^{\alpha}_{a} e^{\rho}_{b}\!=\!\eta_{ab}$ and they are used to pass from coordinate indices to Lorentz indices while the Minkowskian matrix $\eta_{ab}$ is used to move Lorentz (Latin) indices: with it, the Clifford matrices $\boldsymbol{\gamma}^{a}$ are defined as
\begin{eqnarray}
&\{\boldsymbol{\gamma}^{a}\!,\!\boldsymbol{\gamma}^{b}\}\!=\! 2\eta^{ab}\boldsymbol{\mathbb{I}}
\end{eqnarray}
from which
\begin{eqnarray}
&\frac{1}{4}\!\left[\boldsymbol{\gamma}^{a}
\!,\!\boldsymbol{\gamma}^{b}\right]\!=\!\boldsymbol{\sigma}^{ab}
\end{eqnarray}
implicitly defining through
\begin{eqnarray}
&\boldsymbol{\sigma}_{ab}
=-\frac{i}{2}\varepsilon_{abcd}\boldsymbol{\pi}\boldsymbol{\sigma}^{cd}
\end{eqnarray}
the parity-odd matrix $\boldsymbol{\pi}$ used to compute the left-handed and right-handed projections of spinors. The coordinate connection $\Lambda^{\sigma}_{\alpha\nu}$ can be converted into Lorentz connection by means of $\Omega^{a}_{b\pi}\!=\!e^{\nu}_{b}e^{a}_{\sigma} (\Lambda^{\sigma}_{\nu\pi}\!-\!e^{\sigma}_{i}\partial_{\pi}e_{\nu}^{i})$ and known as spin connection: with the gauge potential, it defines
\begin{eqnarray}
&\boldsymbol{\Omega}_{\mu}
=\frac{1}{2}\Omega^{ab}_{\phantom{ab}\mu}\boldsymbol{\sigma}_{ab}
\!+\!iqA_{\mu}\boldsymbol{\mathbb{I}}\label{spinorialconnection}
\end{eqnarray}
as what is said to be spinorial connection, and with which we will define the spinorial covariant derivatives acting on the spinorial fields. In \cite{Fabbri:2014dxa} it has been discussed how it is reasonable to consider the torsion tensor to be completely antisymmetric and in turn this means that it can be written as the dual of an axial vector, which we will indicate according to $W_{\alpha}$ in the following. From the tetrads we may construct the Riemann curvature tensor and its contractions given by the Ricci curvature tensor and scalar, the last indicated with $R$ as usual; then from the gauge potential we may construct the Maxwell strength, indicated with $F_{\alpha\nu}$ as usual; from the torsion axial-vector we may construct the torsion axial-vector curl, which we will indicate with $(\partial W)_{\alpha\nu}$ in the following: these will account for the geometrical quantities. The content of matter is given by the conjugate spinor fields $\overline{\psi}$ and $\psi$ with which
\begin{eqnarray}
&2i\overline{\psi}\boldsymbol{\sigma}^{ab}\psi\!=\!S^{ab}\\
&\overline{\psi}\boldsymbol{\gamma}^{a}\boldsymbol{\pi}\psi\!=\!V^{a}\\
&\overline{\psi}\boldsymbol{\gamma}^{a}\psi\!=\!U^{a}\\
&i\overline{\psi}\boldsymbol{\pi}\psi\!=\!\Theta\\
&\overline{\psi}\psi\!=\!\Phi
\end{eqnarray}
and because $\boldsymbol{\sigma}^{ab}, \boldsymbol{\gamma}^{a}\boldsymbol{\pi}, \boldsymbol{\gamma}^{a}, \boldsymbol{\pi}, \boldsymbol{\mathbb{I}}$ are linearly independent, then these $16$ bi-linear spinor quantities account for all information about the field distribution, although these are actually more than needed: in fact the identity
\begin{eqnarray}
&S_{ab}(\Phi^{2}\!+\!\Theta^{2})\!=\!U^{j}V^{k}\varepsilon_{jkab}\Phi\!+\!U_{[a}V_{b]}\Theta
\end{eqnarray}
tells that the antisymmetric tensor is not necessary whenever one has the vector and axial-vector together with the scalar and pseudo-scalar bilinear; moreover, by writing
\begin{eqnarray}
&V^ {a}\!=\!(\Phi^{2}\!+\!\Theta^{2})^{\frac{1}{2}}v^{a}\label{aux1}\\
&U^{a} \!=\!(\Phi^{2}\!+\!\Theta^{2})^{\frac{1}{2}}u^{a}\label{aux2}
\end{eqnarray}
one obtains the quantities $v^{a}$ and $u^{a}$ verifying
\begin{eqnarray}
&u_{a}u^{a}\!=\!-v_{a}v^{a}\!=\!1\label{norm1}\\
&v_{a}u^{a}\!=\!0\label{orthogonal1}
\end{eqnarray}
telling that vector and axial-vector are orthogonal and of opposite unitary norms; further still, by writing
\begin{eqnarray}
&\Phi\!=\!2\phi^{2}\cos{\beta}\label{b1}\\
&\Theta\!=\!2\phi^{2}\sin{\beta}\label{b2}
\end{eqnarray}
we obtain a re-parametrization of the scalar and pseudo-scalar in terms of two quantities $\phi$ and $\beta$ called module and Takabayashi angle. Consequently, we may conclude that the tensors $R$ and $(\partial W)_{\alpha\nu}$ with $F_{\alpha\nu}$ together with the spinors $\overline{\psi}$ and $\psi$ or equivalently the quantities given by the $v^{a}$ and $u^{a}$ with $\phi$ and $\beta$ contain all information about geometry and its content of matter distributions.

To assign the dynamics, we consider the action to be restricted to the least-order derivative terms (and respecting parity-conservation), whose most general form is
\begin{eqnarray}
\nonumber
&\mathscr{L}\!=\!\frac{1}{4}(\partial W)^{2}\!-\!\frac{1}{2}M^{2}W^{2}
\!+\!\frac{1}{k}R\!+\!\frac{2}{k}\Lambda\!+\!\frac{1}{4}F^{2}-\\
&-i\overline{\psi}\boldsymbol{\gamma}^{\mu}\boldsymbol{\nabla}_{\mu}\psi
\!+\!X\overline{\psi}\boldsymbol{\gamma}^{\mu}\boldsymbol{\pi}\psi W_{\mu}\!+\!m\overline{\psi}\psi
\label{l}
\end{eqnarray}
where $\Lambda$ is the cosmological constant and $M$ and $m$ the masses of torsion and the matter field \cite{Fabbri:2014dxa}: varying with respect to tetrads, torsion, gauge potentials and matter fields, one gets the set of field equations to employ next.

It is important to remark that the positivity of energy and mass terms of torsion is what implies that torsion-matter interactions be attractive: this changes all arguments about the stability conditions discussed in \cite{Fabbri:2015xga}.

We also notice that this general analysis can be further simplified: since the vector bi-linear is time-like we boost into the frame in which it has only its time component, and in this frame we perform rotations bringing the axial-vector bi-linear aligned with the third axis and shifting the phase away: this leaves the spinor in the form
\begin{eqnarray}
&\!\!\psi\!=\!\left(\!\begin{tabular}{c}
$e^{\frac{i}{2}\beta}$\\
$0$\\
$e^{-\frac{i}{2}\beta}$\\
$0$
\end{tabular}\!\right)\!\phi
\label{spinor}
\end{eqnarray}
where $\phi$ and $\beta$ are precisely the module and Takabayashi angle above. The spinor form (\ref{spinor}) clearly shows that the pair of left-handed and right-handed semi-spinor components are complex conjugate of each other; with it we may employ (\ref{spinorialconnection}) to see that the spinor covariant derivative is
\begin{eqnarray}
&\!\!\!\!\!\!\!\!i\boldsymbol{\nabla}_{\mu}\psi
\!=\![(i\nabla_{\mu}\ln{\phi}\!-\!qA_{\mu})\mathbb{I}
\!+\!\frac{1}{2}\nabla_{\mu}\beta\boldsymbol{\pi}
\!+\!\frac{i}{2}\Omega^{ab}_{\phantom{ab}\mu}\boldsymbol{\sigma}_{ab}]\psi\label{scd}
\end{eqnarray}
in the most general decomposition. All these results have been demonstrated and thoroughly discussed in \cite{Fabbri:2016msm}.

By varying the Lagrangian it is possible to get a system of field equations, which can be worked out, by means of expressions (\ref{spinor}, \ref{scd}), in order to be given by the following 
\begin{eqnarray}
\nonumber
&R^{\rho\sigma}\!+\!\Lambda g^{\rho\sigma}\!=\frac{k}{2}[\frac{1}{4}(\partial W)^{2}g^{\rho\sigma}
\!-\!(\partial W)^{\sigma\alpha}(\partial W)^{\rho}_{\phantom{\rho}\alpha}+\\
\nonumber
&+M^{2}W^{\rho}W^{\sigma}\!+\!\frac{1}{4}F^{2}g^{\rho\sigma}
\!-\!F^{\rho\alpha}\!F^{\sigma}_{\phantom{\sigma}\alpha}+\\
\nonumber
&+\frac{1}{4}\phi^{2}(\Omega_{ab}^{\phantom{ab}\rho}\varepsilon^{\sigma abk}v_{k}
\!+\!\Omega_{ab}^{\phantom{ab}\sigma}\varepsilon^{\rho abk}v_{k}-\\
\nonumber
&-\Omega_{ijk}\varepsilon^{ijk\sigma}v^{\rho}
\!-\!\Omega_{ijk}\varepsilon^{ijk\rho}v^{\sigma})-\\
\nonumber
&-q\phi^{2}(A^{\rho}u^{\sigma}\!+\!A^{\sigma}u^{\rho}+\\
\nonumber
&+A_{k}u^{[k}v^{\sigma]}v^{\rho}\!+\!A_{k}u^{[k}v^{\rho]}v^{\sigma})-\\
&\!-2m\phi^{2}\cos{\beta}(\frac{1}{2}g^{\rho\sigma}\!+\!v^{\rho}v^{\sigma})]
\end{eqnarray}
for the gravitational field with
\begin{eqnarray}
&\nabla_{\sigma}F^{\sigma\mu}\!=\!2q\phi^{2}u^{\mu}
\end{eqnarray}
for the gauge field alongside to
\begin{eqnarray}
&\!\nabla_{\alpha}(\partial W)^{\alpha\mu}\!+\!M^{2}W^{\mu}\!=\!2X\phi^{2}v^{\mu}
\end{eqnarray}
for the torsion field, and so that
\begin{eqnarray}
&\!\!\!\!\!\!\!\!\!\!\frac{1}{2}\nabla_{\mu}\beta
\!-\!\!\frac{1}{4}\varepsilon_{\mu\alpha\nu\iota}\Omega^{\alpha\nu\iota}
\!\!+\!qA^{\iota}u_{[\iota}v_{\mu]}\!\!-\!\!XW_{\mu}\!\!+\!v_{\mu}m\cos{\beta}\!=\!0\label{F1}\\
&\!\!\!\!\nabla_{\mu}\ln{\phi}\!-\!\frac{1}{2}\Omega_{\mu a}^{\phantom{\mu a}a}
\!+\!qA^{\rho}u^{\nu}v^{\alpha}\varepsilon_{\mu\rho\nu\alpha}
\!+\!v_{\mu}m\sin{\beta}\!=\!0\label{F2}
\end{eqnarray}
are the spinor field equations. Once again all these results have been demonstrated and discussed in reference \cite{Fabbri:2016laz}.

It is important to notice that in QFT the free particle given by $i\boldsymbol{\nabla}_{\mu}\psi\!=\!P_{\mu}\psi$ is contained in (\ref{scd}) whenever we have $\Omega_{12t}\!=\!2P_{t}\!=\!2m$ with $\nabla_{\mu}\phi\!=\!0$ and $\beta\!=\!0$ which is in fact a non--square-integrable plane-wave solution for the spinor field equations (\ref{F1}, \ref{F2}), as it can be observed.
\section{Spin Precession}
In the previous section, we introduced the general theory we will employ, and next we discuss what features we want our solution to have: first, we think it is reasonable to assume that no gravitation be present; second, because neutrinos carry no charge it is necessary to study systems in which there arises no gauge field. It is not possible to have any guess about what torsion could possibly be, and therefore we will risk no assumption on torsion in general. 

Even without any gravitational field, there may still be some inertial effect interplaying with the spinor field, and so one may choose to have all relevant information either within the connection or within the spinor field: just for example, in the case we have presented here above, the condition given by $\Omega_{12\mu}\!=\!2P_{\mu}$ was due to the fact that the spinor was written as (\ref{spinor}), but alternatively one could also decide to keep this connection equal to zero so long as one is willing to keep a phase $e^{-iP_{\alpha}x^{\alpha}}$ as supplementary multiplicative factor of the spinor (\ref{spinor}); and one may not align the axial-vector bi-linear along the third axis letting the spin axial-vector display a uniform precession around the third axis. Thus, as far as we are concerned, we shall only assume to be in the rest frame of the matter field.

With only the requirement of being in the rest frame, where the spatial component of the velocity vector vanishes, the spinor field in its most general form is
\begin{eqnarray}
&\!\!\psi\!=\!\left(\!\begin{tabular}{c}
$e^{i\upsilon}e^{\frac{i}{2}\beta}e^{-\frac{i}{2}\zeta}\cos{\frac{\xi}{2}}$\\
$e^{i\upsilon}e^{\frac{i}{2}\beta}e^{\frac{i}{2}\zeta}\sin{\frac{\xi}{2}}$\\
$e^{i\upsilon}e^{-\frac{i}{2}\beta}e^{-\frac{i}{2}\zeta}\cos{\frac{\xi}{2}}$\\
$e^{i\upsilon}e^{-\frac{i}{2}\beta}e^{\frac{i}{2}\zeta}\sin{\frac{\xi}{2}}$
\end{tabular}\!\right)\!\phi
\end{eqnarray}
in terms of five independent fields: with this form of the spinor, the spin axial-vector has components
\begin{eqnarray}
&V^{0}\!=\!0\label{0}\\
&V^{1}\!=\!2\phi^{2}\sin{\xi}\cos{\zeta}\label{1}\\
&V^{2}\!=\!2\phi^{2}\sin{\xi}\sin{\zeta}\label{2}\\
&V^{3}\!=\!2\phi^{2}\cos{\xi}\label{3}
\end{eqnarray}
showing that $\xi$ can be taken to be the angle between the spin and the third axis while $\zeta$ is the angle of precession of the spin itself, which consequently can respectively be taken to be constant and $\zeta\!=\!-\omega t$ with $\omega$ being the angular velocity of the precession, and therefore reducing to three the number of independent fields; a final reduction is the one discussed above for which $\upsilon\!=\!-P_{\mu}x^{\mu}$ reducing to two independent fields. Then the spinor has form
\begin{eqnarray}
&\!\!\psi\!=\!\left(\!\begin{tabular}{c}
$e^{i\frac{\beta}{2}}e^{i\frac{\omega}{2}t}\cos{\frac{\xi}{2}}$\\
$e^{i\frac{\beta}{2}}e^{-i\frac{\omega}{2}t}\sin{\frac{\xi}{2}}$\\
$e^{-i\frac{\beta}{2}}e^{i\frac{\omega}{2}t}\cos{\frac{\xi}{2}}$\\
$e^{-i\frac{\beta}{2}}e^{-i\frac{\omega}{2}t}\sin{\frac{\xi}{2}}$
\end{tabular}\!\right)\!e^{-iP_{\mu}x^{\mu}}\!\phi
\label{spinorfield}
\end{eqnarray}
with the two independent fields given by the module and the Takabayashi angle that we have introduced above.

We will work in polar coordinates $(t,\varphi,\theta,r)$ where we have the freedom to choose a pair of dual bases of tetrads
\begin{eqnarray}
&e^{0}_{t}\!=\!1\ \ \ \ 
e^{t}_{0}\!=\!1\\
&e^{1}_{\varphi}\!=\!-r\sin{\theta}\sin{(\omega t)}\ \ \ \  
e^{\varphi}_{1}\!=\!-\frac{1}{r\sin{\theta}}\sin{(\omega t)}\\
&e^{2}_{\varphi}\!=\!-r\sin{\theta}\cos{(\omega t)}\ \ \ \  
e^{\varphi}_{2}\!=\!-\frac{1}{r\sin{\theta}}\cos{(\omega t)}\\
&e^{1}_{\theta}\!=\!r\cos{\xi}\cos{(\omega t)}\ \ \ \  
e^{\theta}_{1}\!=\!\frac{1}{r}\cos{\xi}\cos{(\omega t)}\\
&e^{2}_{\theta}\!=\!-r\cos{\xi}\sin{(\omega t)}\ \ \ \  
e^{\theta}_{2}\!=\!-\frac{1}{r}\cos{\xi}\sin{(\omega t)}\\
&e^{3}_{\theta}\!=\!-r\sin{\xi}\ \ \ \  
e^{\theta}_{3}\!=\!-\frac{1}{r}\sin{\xi}\\
&e^{1}_{r}\!=\!\sin{\xi}\cos{(\omega t)}\ \ \ \  
e^{r}_{1}\!=\!\sin{\xi}\cos{(\omega t)}\\
&e^{2}_{r}\!=\!-\sin{\xi}\sin{(\omega t)}\ \ \ \  
e^{r}_{2}\!=\!-\sin{\xi}\sin{(\omega t)}\\
&e^{3}_{r}\!=\!\cos{\xi}\ \ \ \  
e^{r}_{3}\!=\!\cos{\xi}
\end{eqnarray}
all other components equal to zero; these tetrads give
\begin{eqnarray}
&\Omega_{12t}\!=\!\omega\label{connection1}\\
&\Omega_{12\varphi}\!=\!-\cos{(\xi\!-\!\theta)}\label{connection2a}\\
&\Omega_{23\varphi}\!=\!-\cos{(\omega t)}\sin{(\xi\!-\!\theta)}\\
&\Omega_{31\varphi}\!=\!\sin{(\omega t)}\sin{(\xi\!-\!\theta)}\\
&\Omega_{13\theta}\!=\!-\cos{(\omega t)}\\
&\Omega_{23\theta}\!=\!\sin{(\omega t)}\label{connection2b}
\end{eqnarray}
all other components equal to zero; this spin connection yields a curvature equal to zero: this is expected because we have assumed the absence of the gravitational field.

Such a choice of tetrad fields has been made to ensure that they rotate following the uniform precession of the spin axial-vector and therefore it should be expected that with them (\ref{0}, \ref{1}, \ref{2}, \ref{3}) reduce to
\begin{eqnarray}
&V^{t}\!=\!0\\
&V^{\varphi}\!=\!0\\
&V^{\theta}\!=\!0\\
&V^{r}\!=\!2\phi^{2}
\end{eqnarray}
which is in fact quite considerably simplified.

With the spinorial connection we may compute
\begin{eqnarray}
\nonumber
&i\boldsymbol{\nabla}_{\mu}\psi\!=\![\frac{1}{2}\nabla_{\mu}\beta\boldsymbol{\pi}
\!+\!i(\frac{1}{2}\Omega^{ab}_{\phantom{ab}\mu}\boldsymbol{\sigma}_{ab}
\!-\!\omega\partial_{\mu}t\boldsymbol{\sigma}_{12})+\\
&+(P_{\mu}\!+\!i\nabla_{\mu}\ln{\phi})\mathbb{I}]\psi\label{scdf}
\end{eqnarray}
which tells that for $\mu\!=\!t$ the rotation of the tetrads and the precession of the spin axial-vector cancel one another, as expected from the considerations above; for all spatial coordinates there remains a residual contribution in the spin connection accounting for the inertial effects of the polar coordinates. Employing (\ref{connection1}-\ref{connection2b}) into (\ref{scdf}), and then following the same method presented in \cite{Fabbri:2016laz}, we obtain the field equations in the form we will use in the following.

The torsion field equations are given by
\begin{eqnarray}
&\!\nabla_{\alpha}(\partial W)^{\alpha\mu}\!+\!M^{2}W^{\mu}\!=\!2X\phi^{2}v^{\mu}
\end{eqnarray}
and the spinorial matter field equations are
\begin{eqnarray}
&\frac{1}{2}\nabla_{\mu}\beta\!-\!P^{\iota}u_{[\iota}v_{\mu]}\!-\!XW_{\mu}
\!+\!v_{\mu}m\cos{\beta}\!=\!0\\
&\frac{1}{2}\nabla_{\mu}\ln{(\phi^{2}r^{2}\sin{\theta})}
\!-\!P^{\rho}u^{\nu}v^{\alpha}\varepsilon_{\mu\rho\nu\alpha}\!+\!v_{\mu}m\sin{\beta}\!=\!0
\end{eqnarray}
where torsion is determined in terms of the module, while the module is determined in terms of the Takabayashi angle and the Takabayashi angle is determined in terms of torsion: the three equations constitute a closed system.

From these field equations it follows the validity of the partially conserved axial current given in the form
\begin{eqnarray}
&M^{2}\nabla_{\mu}W^{\mu}\!=\!4Xm\phi^{2}\sin{\beta}
\label{pcac}
\end{eqnarray}
which will be useful in the following of the study.
\section{Localization}
To begin our analysis of the exact solutions and, more specifically, their character, we consider the momentum
\begin{eqnarray}
&P_{\mu}\!=\!(E,0,0,0)\label{momentum}
\end{eqnarray}
where $E$ is the energy of the spinor, and we specify that in choosing this condition our goal is simply that we wish to obtain solutions that are close to the standard solutions one would have in QFT: like in QFT and, more generally, in any theory in which gravity is neglected, the energy is a free parameter, but because of the presence of torsion, which is attractive, the potential is a well, thus negative, so that the energy is smaller than the mass, and thus the condition $m\!>\!E$ may be fixed. Since torsion is massive, we take, for small momenta, the effective approximation, in which the torsion field equations are
\begin{eqnarray}
&M^{2}W_{\mu}\!=\!2X\phi^{2}v_{\mu}\label{approximation}
\end{eqnarray}
so that torsion can be integrated in the spinorial matter field equations: notice that this approximation does not spoil the validity of the partially conserved axial current, which can then still be used to show that, for attractive interactions, the Takabayashi angle is negative, and thus condition $\beta\!<\!0$ can be taken. We will study situations in which the Takabayashi angle can be taken to be small.

The spinorial matter field equations reduce to the form
\begin{eqnarray}
&\nabla_{\mu}\beta\!-\!m\beta^{2}v_{\mu}\!+\!2|m\!-\!E|v_{\mu}
\!-\!4\frac{X^{2}}{M^{2}}\phi^{2}v_{\mu}\!=\!0\label{f1}\\
&\nabla_{\mu}\ln{(\phi^{2}r^{2}\sin{\theta})}\!+\!2v_{\mu}m\beta\!=\!0\label{f2}
\end{eqnarray}
where again there are two field equations for two independent fields: the system of field equations is still closed.

By taking into account the expression given above for the normalized axial-vector it is possible to see that the field equations (\ref{f1}, \ref{f2}) are explicitly given by
\begin{eqnarray}
&\partial_{r}\beta\!+\!m\beta^{2}\!-\!2|m\!-\!E|\!+\!4\frac{X^{2}}{M^{2}}\phi^{2}
\!\approx\!0\label{zv}\\
&\partial_{r}\ln{(\phi^{2}r^{2}\sin{\theta})}\!-\!2m\beta\!\approx\!0
\label{zf}
\end{eqnarray}
in which we can clearly see that the temporal dependence has disappeared, henceforth showing that solutions are stationary and therefore stable, beyond the fact that also the azimuthal dependence has disappeared, thus showing that the solutions are axially symmetric, as expected.

The field equations (\ref{zv}, \ref{zf}) can be combined yielding the second-order field equation for the module in the form
\begin{eqnarray}
\nonumber
&\partial_{r}\partial_{r}(\phi r\sqrt{\sin{\theta}})
\!+\!\frac{4X^{2}m}{M^{2}r^{2}\sin{\theta}}(\phi r\sqrt{\sin{\theta}})^{3}-\\
&-2m|m\!-\!E|(\phi r\sqrt{\sin{\theta}})\!\approx\!0
\end{eqnarray}
in which we see the presence of an attractive interaction and a negative-sign mass-like term in what can be viewed as a spherical analog of a soliton field equation: solutions in general are not known because of the cubic term, but whichever solution it may have, such solution has to drop toward infinity, so for large-$r$ regions the cubic term tends to vanish and hence the above field equation reduces to
\begin{eqnarray}
&\partial_{r}\partial_{r}(\phi r\sqrt{\sin{\theta}})
\!-\!2m|m\!-\!E|(\phi r\sqrt{\sin{\theta}})\!\approx\!0
\end{eqnarray}
which as a matter of fact has quite well known solutions.

Such solutions are given by the following form
\begin{eqnarray}
&\phi r\sqrt{\sin{\theta}}\!\approx\!K\left[\exp{\left(r\sqrt{2m|m-E|}\right)}\right]^{-1}
\end{eqnarray}
for any given constant $K$ as an exponential damping with the distance, therefore displaying a drop toward infinity, and in fact such a drop toward infinity is so fast that its volume integral is finite, and the distribution is localized.

With this solution for the module we may employ field equation (\ref{zf}) to get the Takabayashi angle
\begin{eqnarray}
&\frac{\beta}{2}\!\approx\!-\sqrt{\frac{|m-E|}{2m}}
\end{eqnarray}
in which we dropped the irrelevant integration constant and which can be plugged into field equation (\ref{zv}) to show that the consistency of both matter field equations checks straightforwardly within the assumed approximations.

We also remark that equations (\ref{approximation}) show that torsion is treated effectively giving rise to a model that is essentially the Nambu--Jona-Lasinio model, where the interaction is known to be attractive and yielding bound states.
\section{Conclusion}
In this paper, we have considered the theory of torsion gravity in the case of the Dirac field: then we have been studying what happens when the spinor field is described by a general matter distribution in the frame in which its spatial velocity is equal to zero while the spatial spin possesses a constant angle with the third axis and a uniform precession around it; we have chosen the tetrad fields rotating in such a manner as to follow such precession, but no gravity nor any other interaction has been taken into account beside the torsional contribution. A momentum was chosen, torsion was taken in effective approximation, and the Takabayashi angle was taken negative although very small. The large-distance behaviour of the module was studied, and we found that it consisted in decreasing exponentials, so all relevant quantities were found to be localized, and exact solutions would be square-integrable.

There are four factors ensuring stable and localized solutions: the mass is necessary in order to avoid a constant behaviour; an energy smaller than the mass is necessary to avoid non-constant oscillatory behaviour; a small and negative Takabayashi angle gives non-oscillatory decreasing behaviour; non-zero axial-vector components correspond to coordinates with respect to which the decreasing behaviour of matter distributions occurs. If the tensor of torsion is coupled to the spin-dependent bi-linear quantities given by the spin axial-vector and the pseudo-scalar, a negative Takabayashi angle and an energy smaller than the mass are not only conceivable but also reasonable.

The fact that torsion could give rise to localized matter distributions is expected because the coupling between an axial-vector massive boson and the spin gives attractive interactions, and if they are strong enough then they can provide the conditions for the existence of bound states as it is well known from the Nambu--Jona-Lasinio model.

Localized distributions are a general character resulting from the presence of attractive interactions, and they commonly come in the form of soliton solutions when the attractive interaction is in the form of a non-linear potential: non-linear equations admitting soliton-like solutions have been studied for example in \cite{t}. Another source of reference for solitonic behaviour can be found in \cite{MacKenzie:2001av}.

The type of non-linear potential that gives rise to such attractive interactions is quite generally obtained from a theory in which torsion is in effective approximation.

In order for this not to be so one would need to require the torsion mass not to be large enough: then one might well wonder what would happen if the torsion were to be considered in the most general propagating case.

However, in even more extended theories torsion has higher-order terms violating renormalizability.

Instead some possible as well as interesting effect could arise from non-trivial gravitational fields.

\end{document}